\documentstyle[aps,epsf,multicol]{revtex}
\newcommand{\braket}[1]{\langle #1 \rangle}
\def \be {\begin{equation}}
\def \ee {\end{equation}}
\def \bea {\begin{eqnarray}}
\def \eea {\end{eqnarray}}
\def \bfi {\begin{figure}}
\def \efi {\end{figure}}

\begin{document}

\title{Bethe Ansatz Solutions and 
	Excitation Gap of the Attractive Bose-Hubbard Model} 
\author{Deok-Sun Lee and Doochul Kim}
\address{
School of Physics, Seoul National University, 
Seoul 151-747, Korea}

\maketitle

\begin{abstract}
The energy gap between the ground state and the first excited 
state of the one-dimensional attractive Bose-Hubbard Hamiltonian 
is investigated in connection with directed polymers in 
random media.
The excitation gap $\Delta$ is obtained by exact diagonalization
of the Hamiltonian in the two- and three-particle sectors and 
also by an exact Bethe Ansatz solution in the two-particle sector.
The dynamic exponent $z$ is found to be $2$. However, in the 
intermediate range of the size $L$ where $UL\sim {\cal O}(1)$,
$U$ being the attractive interaction, the effective dynamic 
exponent shows an anomalous peak reaching high values of $2.4$
and $2.7$ for the two- and the three-particle sectors, respectively.
The anomalous behavior is related to a change in the sign of
the first excited-state energy. In the two-particle sector, we
use the Bethe Ansatz solution to obtain the effective dynamic
exponent as a function of the scaling variable $UL/\pi$.
The continuum version, the attractive delta-function Bose-gas 
Hamiltonian, is integrable by the Bethe Ansatz with suitable 
quantum numbers, the distributions of which are not known in 
general. Quantum numbers are proposed for the first excited state 
and are confirmed numerically for an arbitrary number of 
particles.
\end{abstract}

\begin{multicols}{2}

\section{Introduction}

The dynamics of many simple non-equilibrium systems are often 
studied through corresponding quantum Hamiltonians. Examples
are the asymmetric $XXZ$ chain Hamiltonian and the attractive
Bose-Hubbard Hamiltonian for the single-step growth 
model~\cite{plischke87} and the directed polymers in random
media (DPRM)~\cite{kardar87}, respectively. 
The single-step growth model is
a Kardar-Parisi-Zhang (KPZ) universality class growth model where
the interface height $h(x,t)$ grows in a stochastic manner
under the condition that $h(x\pm1,t)-h(x,t)=\pm 1$. The process
is also called the asymmetric exclusion process (ASEP) in a different
context. The evolution of the probability distribution for $h(x,t)$ 
is generated by the asymmetric $XXZ$ chain Hamiltonian~\cite{spohn92}. 
The entire information about the dynamics is coded in 
the generating function $\braket{e^{\alpha h(x,t)}}$. Its time
evolution, in turn, is given by the modified asymmetric $XXZ$ chain
Hamiltonian~\cite{derrida98,dslee99,appert99},
\be
H_{\rm XXZ}(\alpha)=
-\sum_{i=1}^L \left\{
e^{2\alpha/L}\sigma_i^-\sigma_{i+1}^++
{1\over 4}(\sigma_i^z \sigma_{i+1}^z-1)\right\},
\label{XXZ}
\ee
with the $\sigma$'s being the Pauli matrices and 
$L$ the size of the system.
(We consider only the periodic boundary condition in this work.)
On the other hand, in the DPRM problem, the partition function 
$Z(x,t)$, the partition sum over directed polymer configurations
with fixed ends at $(0,0)$ and $(x,t)$, is the quantity of main
interest. Its generating function $\braket{{Z(x,t)}^n}$ then 
evolves by the one-dimensional attractive Bose-Hubbard 
Hamiltonian~\cite{kardar87},
\bea
H_{\rm BH}(n)&=&-{1\over 2}\sum_{i=1}^L(b_i b_{i+1}^\dagger+ 
b_i^\dagger b_{i+1}-2)\nonumber \\ 
&&-
U\sum_{i=1}^L{b_i^\dagger b_i(b_i^\dagger b_i-1)\over 2}.
\label{BH}
\eea
Here, the $b^\dagger$'s($b$'s) are the boson creation (annihilation)  
operators, $\sum_{i=1}^Lb_i^\dagger b_i=n$ is the conserved particle
number, and $U(>0)$ is the attractive interaction.
The two systems are closely related, at the level of continuum 
stochastic differential equations, through the Cole-Hopf 
transformation, $Z(x,t)=e^{-h(x,t)}$~\cite{krug91}.
In particular, $U$ in Eq.~(\ref{BH}) is
related to the particle density, $\rho=\sum_{i=1}^L (\sigma_i^z +1)/2$ 
of Eq.~(\ref{XXZ}), by $U=4\rho(1-\rho)$~\cite{derrida97,park01}. 
Recently, the universal probability distribution function (PDF) 
at the stationary state was found for both models mentioned 
above~\cite{derrida98,dslee99,appert99,brunet00}. 
Their common structure of the ground-state energies 
as functions of the scaling variables 
$\alpha\sqrt{4L\rho(1-\rho)}$ and $-n\sqrt{LU}$, 
for the single-step growth model and the DPRM problem, 
respectively,  
gives the universal PDF at the stationary state.

While the ground-state energy of each Hamiltonian gives 
information about the stationary state of the corresponding
process, 
the first excited-state energy, combined with the ground-state 
energy, is related to the characteristic behavior of the process 
as it approaches the stationary state. 
For example, in the single-step growth model, as $t\to \infty$, 
the generating function takes the form 
\be
\log \braket{e^{\alpha h(x,t)}} \sim 
\left\{-E_0(\alpha)t + e^{-\Delta(\alpha) t}\right\},
\ee  
and similarly for $\log\braket{{Z(x,t)}^n}$.
Here, $E_0(\alpha)$ is the ground-state energy of Eq.~(\ref{XXZ}), 
$\Delta(\alpha)$ is the inverse of the relaxation time, as well as  
the gap between the ground-state energy and the first
excited-state energy, $E_1(\alpha)$, such that 
$\Delta(\alpha)=E_1(\alpha)-E_0(\alpha)$.
The size dependence of $\Delta(\alpha)$, $\Delta(\alpha)\sim L^{-z}$, 
defines the dynamic exponent $z$.

Because the asymmetric $XXZ$ chain Hamiltonian, 
$H_{\rm XXZ}(\alpha)$, is integrable by the Bethe 
Ansatz, the low-lying state energies, as well as 
the size dependence of the excitation gap, are well understood.  
When $\alpha\sqrt{4L\rho(1-\rho)}\gg 1$ and the density of 
particles is finite in the limit $L\to\infty$,
$\Delta(\alpha)$ behaves as $\Delta(\alpha)\sim L^{-1}$.
However, when $\alpha\sqrt{4L\rho(1-\rho)}\sim {\cal O}(1)$, 
$\Delta(\alpha)$ behaves as 
$\Delta(\alpha) \sim L^{-3/2}$~\cite{spohn92,dkim95}.
The dynamic exponent $z=3/2$ is a characteristic of the 
dynamic universality class of the KPZ-type surface growth. 
When the number of particles is finite and the density of
particles is very low, it is known that 
$\Delta(\alpha)\sim L^{-2}$~\cite{henkel94}.
However, when $\alpha<0$, which corresponds to the ferromagnetic
phase, most Bethe Ansatz solutions are not available although
the Bethe Ansatz equations continue to hold. 
As $\alpha$ becomes negative, the quasi-particle momenta appearing in  
the Bethe Ansatz equations become complex, 
so solutions are difficult to obtain analytically. 
  
The attractive Bose-Hubbard Hamiltonian is expected to have some 
resemblance to the ferromagnetic phase of the asymmetric $XXZ$ chain
Hamiltonian considering the equivalence of $\alpha$ and $-n$. 
The equivalence is identified indirectly by comparing the two scaling 
variables $\alpha\sqrt{4L\rho(1-\rho)}$ and $-n\sqrt{LU}$ 
under the relation $U=4\rho(1-\rho)$ or the two generating functions  
$\braket{\exp(\alpha h(x,t)}$ and  $\braket{{Z(x,t)}^n}$ under the 
relation $Z(x,t)=e^{-h(x,t)}$. 
In contrast to the asymmetric $XXZ$ chain Hamiltonian,
the Bose-Hubbard Hamiltonian does not satisfy the Bethe Ansatz except 
in the two-particle sector~\cite{haldane82}. 
Instead, the attractive delta-function Bose-gas Hamiltonian,
\be
H_{\rm D}(n)= -{1\over 2}\sum_{i=1}^n {\partial^2 \over \partial x_i^2} -
U \sum_{i<j} \delta(x_i-x_j), 
\label{DELTA}
\ee
which is the continuum version of the attractive Bose-Hubbard 
Hamiltonian,
is known to be integrable by the Bethe Ansatz. 
The attractive delta-function Bose gas has been studied in 
Refs.~\cite{lieb63} and~\cite{Muga98}.
The ground-state energy is obtained from the 
Bethe Ansatz solution by using the symmetric distribution of the 
purely imaginary quasi-particle momenta. 
However, the structure of the energy spectra is not well known for the
same reason as in the asymmetric $XXZ$ chain Hamiltonian with
$\alpha<0$. 
The unknown energy spectra itself prevents one from understanding the
dynamics of DPRM near the stationary state. 

In this paper, we discuss in Section II the distribution of the 
quantum numbers appearing in the Bethe Ansatz equation for the 
first excited state of the attractive delta-function Bose-gas 
Hamiltonian, the knowledge of which is essential for 
solving the Bethe Ansatz equation.
In Section III, the excitation gap of the attractive Bose-Hubbard 
Hamiltonian with a small number of particles is 
investigated through the exact diagonalization method. 
We show that the gap decays as 
$\Delta\sim L^{-2}$, i.e., $z=2$, 
but that the exponent becomes anomalous when 
$U\sim L^{-1}$.
The emergence of the anomalous exponent is explained in connection 
with the transition of the first excited state from a positive 
energy state to a negative energy state. 
The Bethe Ansatz solutions in the two-particle sector show how 
the behavior of the gap varies with the interaction.
We give a summary and discussion in Section IV.
   
\section{Quantum number distribution for the first excited state}

In this section, we study the Bethe Ansatz solutions for the 
ground state and the first excited state of the attractive delta-function 
Bose-gas Hamiltonian. 
The eigenstate of $H_{\rm D}(n)$, Eq.~(\ref{DELTA}), is of the form 
\bea
\phi(&&x_1,x_2,\ldots,x_n)\nonumber \\
&&=\sum_P
A(P)\exp(ik_{P1}x_1+ik_{P2}x_2+\cdots+ik_{Pn}x_n),
\eea
where $P$ is a permutation of ${1,2,\ldots,n}$ and $x_1\le x_2\le
\ldots \le x_n$ with no three $x$'s being equal. 
The quasi-particle momenta $k_j$'s are determined by solving the Bethe
Ansatz equations,
\bea
&&k_jL=2\pi I_j + \sum_{l\neq j}\theta({k_j-k_l \over
-U}) \ \ (j=1,2,\ldots,n), \nonumber \\ 
&&\theta(x)=-2\tan^{-1}(x).
\label{BAE}
\eea
If the distribution of quantum numbers $\{I\}$ is given, 
the set of quasi-particle momenta $\{k\}$ is uniquely determined.
With such $k_j$'s, the energy eigenvalue is simply given by 
$E=\sum_{j=1}^n (k_j^2/2)$.

For the ground state, the set of quantum numbers $\{I\}$ is
\be
I_j=-{n+1\over 2}+j, \ \ (j=1,2,\ldots,n),
\ee
and the quasi-particle momenta are distributed symmetrically on
the imaginary axis in the complex-$k$ plane. 
Care should be taken when dealing with the first excited state. 
For the repulsive
delta-function Bose-gas Hamiltonian, where $U$ is replaced by $-U$ in
Eq.~(\ref{DELTA}), the quantum numbers for one of the first excited 
states are
\bea
I_j&=&-{n+1\over 2}+j \ \ (j=1,2,\ldots,n-1), \nonumber \\
I_n&=&{n+1\over 2}.
\eea
However, for the attractive case, by following the movement of the
momenta as $U$ changes sign, we find that the quantum numbers for the 
first excited state should be given by
\bea
I_j&=&-{n-1\over 2}+j \ \ (j=1,2,\ldots,n-1), \nonumber \\
I_n&=&-{n-3\over 2}\ (=I_1). 
\eea
That is, the two quantum numbers $I_1$ and $I_n$ become the same.
Such a peculiar distribution of $I_j$'s does not appear in 
other Bethe Ansatz solutions such as those for the $XXZ$ chain 
Hamiltonian or the repulsive delta-function Bose-gas Hamiltonian.
We remark that even though the two $I_j$'s are the same, all
$k_j$'s are distinct; otherwise, the wavefunction vanishes.
Such a distribution of quantum numbers is confirmed by the 
consistency between the energies obtained by diagonalizing the 
Bose-Hubbard Hamiltonian exactly and those obtained by solving 
the Bethe Ansatz equations with the above quantum numbers for 
very weak interactions, for which the two Hamiltonians 
possess almost the same energy spectra.

When there is no interaction ($U=0$), all quasi-particle momenta,  
$k_j$'s, are zero for the ground state while  for the first excited
state, all the $k_j$'s are zero except the last one, $k_n=2\pi/L$.
In the complex-$k$ plane, as the very weak repulsive interaction is 
turned on, the $n-1$ momenta are shifted infinitesimally from $k=0$ 
with $k_1<k_2<\cdots<k_{n-1}$, and the $n$th momentum is shifted 
infinitesimally to the left from $k=2\pi/L$.  All the momenta remain 
on the real axis.
When the interaction is weakly attractive, the $n-1$ momenta 
become complex with 
${\rm Im} \ k_1<{\rm Im} \ k_2<\cdots<{\rm Im} \ k_{n-1}$ and 
${\rm Re} \ k_j \simeq 0$ for $j=1, 2,\ldots, n-1$, and
the $n$th momentum remains on the real axis, 
but is shifted to the left. 
Figure~\ref{momenta} shows the distribution of the quantum 
numbers and the quasi-particle momenta in the presence of 
a very weak attractive interaction. The quasi-particle momenta 
are obtained by solving Eq.~(\ref{BAE}). 

Knowledge of the distribution of the quantum numbers 
is essential for solving the Bethe Ansatz equations of the
attractive delta-function Bose-gas Hamiltonian. 
For the original attractive Bose-Hubbard Hamiltonian, 
the Bethe Ansatz solutions are the exact solutions for the 
two-particle sector only, but are good approximate solutions
in other sectors  provided the density is 
very low and the interaction is very weak. 
This is because the Bethe Ansatz for the Bose-Hubbard Hamiltonian
fails once states with sites occupied by more than three particles
are included. Thus, for the sectors with three or more particles,
the Bethe Ansatz solutions may be regarded as approximate eigenstates
provided states with more than three particles at a site do not play 
an important role in the eigenfunctions.
In Ref.~\cite{haldane82}, 
it is shown that the error in the  Bethe Ansatz
due to multiply-occupied sites (occupied by more than three 
particles) is proportional to $U^2$, where $U(>0)$ 
in Ref.~\cite{haldane82} 
corresponds to  $-U$ in Eq.~(\ref{BH}). 
This applies to the attractive interaction case also. 
For the repulsive Bose-Hubbard Hamiltonian, the Bethe
Ansatz is a good approximation when the density is low
and the interaction is strong because the strong repulsion
prevents many particles from occupying the same site~\cite{krauth91}.
For the attractive Bose-Hubbard Hamiltonian, the Bethe Ansatz is good
when the density is low and the interaction is weak because
a weak attraction is better for preventing many particles
from occupying the same site and because the error is proportional
to $U^2$.

\section{power-law dependence and anomalous exponent}

We are interested in the scaling limit $L\to\infty$ with the 
scaling variable $n\sqrt{UL}$ fixed because  
the common structure of the ground-state energies of 
the asymmetric $XXZ$ chain Hamiltonian and 
of the attractive Bose-Hubbard Hamiltonian is found 
in this scaling limit.
In this section, we investigate the size dependence of the 
excitation gap by using exact diagonalization of the attractive 
Bose-Hubbard Hamiltonian in the two- and the three-particle 
sectors for $L$ up to $30$. Also, for the two-particle sector,
we solve the Bethe Ansatz equation using the result of the previous
section for larger $L$.

Figure~\ref{energy} shows the ground-state energies and the first 
excited-state energies versus the size of the system in the two-  
and the three-particle sectors for three values of $U$.  
Note that for some value of $U$, 
the sign of the first excited-state 
energy changes as the size of the system increases 
while for other values of $U$, no such crossover is seen in 
the range of $L$ investigated here.

The excitation gaps versus the size of the system are shown in 
Fig.~\ref{n=2gap} and Fig.~\ref{n=3gap} on a logarithmic scale. 
For $U=0.05$ and $5$, the nearly straight lines indicate
the power-law behavior of the gap, $\Delta \sim L^{-z}$, and 
the slopes of the fitted lines indicate $z\simeq 2.0$. 
However, for $U=0.5$, the asymptotic behavior shows up only after
a large crossover region where the effective $z$, $z_{\rm eff}$, 
is anomalously large.
For the two-particle sector with $U=0.5$, 
$z_{\rm eff}$ is about $2.4$ in the range $14\leq L \leq 18$. 
For the three-particle sector with $U=0.5$, 
$z_{\rm eff}$ is about $2.7$ in the corresponding range 
$8\leq L \leq 12$.

Looking into the first excited-state energy in Fig.~\ref{energy}, 
one can see that the anomalous value of $z_{\rm eff}$ appears in the 
range of $L$ where the first excited-state energies change their
signs. 
We conjecture that the wavefunction has a transition
as the sign of the energy changes.
In order to confirm the connection between the anomalous exponent 
and the transition of the first excited state, we solved the 
Bethe Ansatz equation to evaluate the effective exponent $z_{\rm eff}$ 
near the transition point in the two-particle sector. 

In the two-particle sector, the quantum numbers for the ground state
and the first excited state are $\{-1/2, 1/2\}$ and
$\{1/2,1/2\} \ ({\rm or} \ \{-1/2,-1/2\})$,
respectively.
The quasi-particle momenta are purely imaginary, i.e., 
$k_1 = - i\kappa$ and $k_2 = i\kappa$ for the ground state. 
For the first excited state, as noted already in Ref.~\cite{lieb63},
when $U<(4/L)\cos (\pi/L)$, the two quasi-particle momenta are real, 
i.e., $k_1={\pi/ L}-k$ and $k_2={\pi/ L}+k$, while when 
$U>(4/L)\cos (\pi/L)$, $k$ is replaced by $iq$ such that 
$k_1=\pi/L-iq$ and $k_2=\pi/L+iq$.
The examples of these distributions are shown in Figs.~\ref{root}
(a) and (b) with $L=100$ 
for the two values of $U=0.001$ and $0.1$, respectively.
The transition of the first excited state 
occurs when the quasi-particle momenta are 
imaginary.  
When $U>(4/L)\cos (\pi/L)$, the ground-state energy($E_0$) 
and the first excited-state energy($E_1$) are, respectively,
given by
\be
E_0=-4 \sinh^2 \left({\kappa\over 2}\right), 
\label{energy_n=2_g}
\ee
and
\be
E_1=4 \sin^2 \left({\pi\over 2L}\right) \cosh^2 
\left({q\over 2}\right) -
4 \cos^2 \left({\pi\over 2L}\right) \sinh^2 \left({q\over 2}\right),
\label{energy_n=2_e}
\ee
where $\kappa$ and $q$ are real and satisfy
\be
\kappa L=\log\left({2\sinh\kappa+U \over 2\sinh\kappa-U}\right), 
\label{BAE_n=2_g}
\ee
and
\be
qL=\log\left({U+2\cos({\pi\over L})\sinh q \over 
U-2\cos({\pi\over L})\sinh q}\right),
\label{BAE_n=2_e}
\ee 
respectively. 
We now consider the scaling limit $L\to\infty$ with $UL$ finite.
Let $U=U^*\equiv(\pi/L)s_U$ 
with $s_U\equiv 2\coth(\pi/2)\simeq 2.181$. At this value of $U$, 
the first excited-state energy, $E_1$,
is $0$, $\kappa^*\equiv\kappa(U^*)= (\pi/L)s_{\kappa}$, 
and $q^*\equiv q(U^*)=\pi/L$. Here, $s_{\kappa}$ satisfies
\be
\pi s_{\kappa}=\log\left({2 s_{\kappa}+s_U \over 2 s_{\kappa}-s_U}\right),
\label{kappa}
\ee
which gives $s_{\kappa} \simeq 1.151$. 
When the size of the system $L$ is increased by $\delta L$ with $U=U^*$,
the changes of $\kappa$ and $q$, $\delta\kappa$ and $\delta q$, are,
from Eqs.~(\ref{BAE_n=2_g}) and 
(\ref{BAE_n=2_e}),
\bea
\delta\kappa&=&-{\pi s_{\kappa}(4{s_{\kappa}}^2-{s_U}^2) 
\over 4{s_{\kappa}}^2-{s_U}^2 +(4/\pi)s_U} \ 
{\delta L\over L^2} \equiv 
-\pi \Gamma \ {\delta L\over L^2}, \nonumber \\
\delta q &=&  
{\pi ({s_U}^2-4) \over (4/\pi)s_U-{s_U}^2+4}
\ {\delta L\over L^2} \equiv
\pi\Sigma \ {\delta L\over L^2}.
\eea
The perturbative expansion $\Delta(L+\delta L) \simeq
\Delta(L)(1-z(\delta L/L))$, 
under the assumption that $\Delta(L)\sim L^{-z}$, gives the
value of $z_{\rm eff}$ at $U^*$:
\be
z_{\rm eff}=2 {1+s_{\kappa}\Gamma+\Sigma \over s_{\kappa}^2}.
\ee
Numerical solutions of Eq.~(\ref{kappa}) give 
$z_{\rm eff}\simeq 2.401$.

On the other hand, when $U\ll U^*$, the ground-state energy, 
$E_0$, is ${\cal O}(U/L)$, and the first excited-state energy, 
$E_1$, is ${\cal O}(1/L^2)$, which are easily obtained from 
Eqs.~(\ref{energy_n=2_g}) and (\ref{energy_n=2_e}). 
Therefore, the excitation gap behaves as $\Delta\sim L^{-2}$.     
Also, when $U^*\ll U\ll 1$, the quasi-particle momenta of the 
ground state are $\pm i \sinh^{-1}(U/2) -{\cal O}(e^{-L})$,  
and those of the first excited state are 
$\pi/L\pm i\sinh^{-1}(U/2) +{\cal O}(e^{-L})$, which lead to 
$z_{\rm eff}=2$.

For arbitrary $U$, the effective value of $z_{\rm eff}$ is 
evaluated from the relation
\be
z_{\rm eff}={\log(\Delta(L+1)/\Delta(L-1))\over \log((L-1)/(L+1))}
\ee
by using the solutions of Eqs.~(\ref{BAE_n=2_g}) and 
(\ref{BAE_n=2_e}) for sufficiently large $L$.
As discussed above, the exponent $z_{\rm eff}$ shows an 
anomalous peak 
near $U=U^*$ or $UL/\pi=s_U$ and approaches $2.0$ as 
$UL/\pi\to 0$ or $\infty$.
Figure~\ref{z} shows a plot of $z_{\rm eff}$ versus the scaling
variable $UL/\pi$ at $L=10000$. 

\section{Summary and Discussion} 

As the asymmetric $XXZ$ chain generates the dynamics of the 
single-step growth model, the attractive Bose-Hubbard Hamiltonian
governs the dynamics of the DPRM. 
We studied the attractive Bose-Hubbard Hamiltonian and its 
continuum version, the attractive delta-function Bose-gas Hamiltonian 
concentrating on the behavior of the excitation gap, which 
is related to the characteristics of DPRM relaxing into the 
stationary state. 
For the attractive delta-function Bose gas Hamiltonian, 
The quantum numbers for the first excited state 
in the Bethe Ansatz equation are found 
for the attractive delta-function Bose gas Hamiltonian, 
and the distribution of the quasi-particle momenta 
is discussed in the presence of a very weak attractive interaction.
Our result is the starting point for a further elucidation of 
the Bethe Ansatz solutions.
We show that the excitation gap depends on the size of the system 
as a power law, $\Delta\sim L^{-z}$, and that the exponent $z$ can be 
calculated 
by using an exact diagonalization of the attractive Bose-Hubbard 
Hamiltonian in the two- and the three-particle sectors 
and by using the Bethe Ansatz 
solution in the two-particle sector. 
The exponent $z$ is $2.0$. However, for the intermediate region where
$UL\sim {\cal O}(1)$, the effective exponent $z_{\rm eff}$ shows a peak.

The equivalence of the differential equations governing the 
single-step growth model and DPRM implies some inherent 
equivalence in the corresponding Hamiltonians.
The power-law behavior of the excitation gap, $\Delta\sim L^{-2}$,
for the attractive Bose-Hubbard Hamiltonian with a very weak 
interaction is the same as that for the asymmetric $XXZ$ chain 
Hamiltonian with a small number of particles, which is expected 
considering the relation $U=4\rho(1-\rho)$.
The fact that the excitation 
gap behaves anomalously for $U\sim L^{-1}$ implies the possibility 
of an anomalous dynamic exponent $z$ for a finite scaling variable 
$n\sqrt{UL}$. If that is the case, one may expect the existence
of some singularity at $n\sim L^{-1/2}$ with finite $U$, where the 
dynamic exponent $z$ takes an anomalous value, which is similar to the 
singularity at $\alpha\sim L^{-1/2}$ for the asymmetric $XXZ$ chain 
Hamiltonian.  

\acknowledgments
This work was supported by the Korea Research Foundation (2000-
015-DP0138).

\bfi
\centerline{\epsfxsize=9cm  \epsfbox{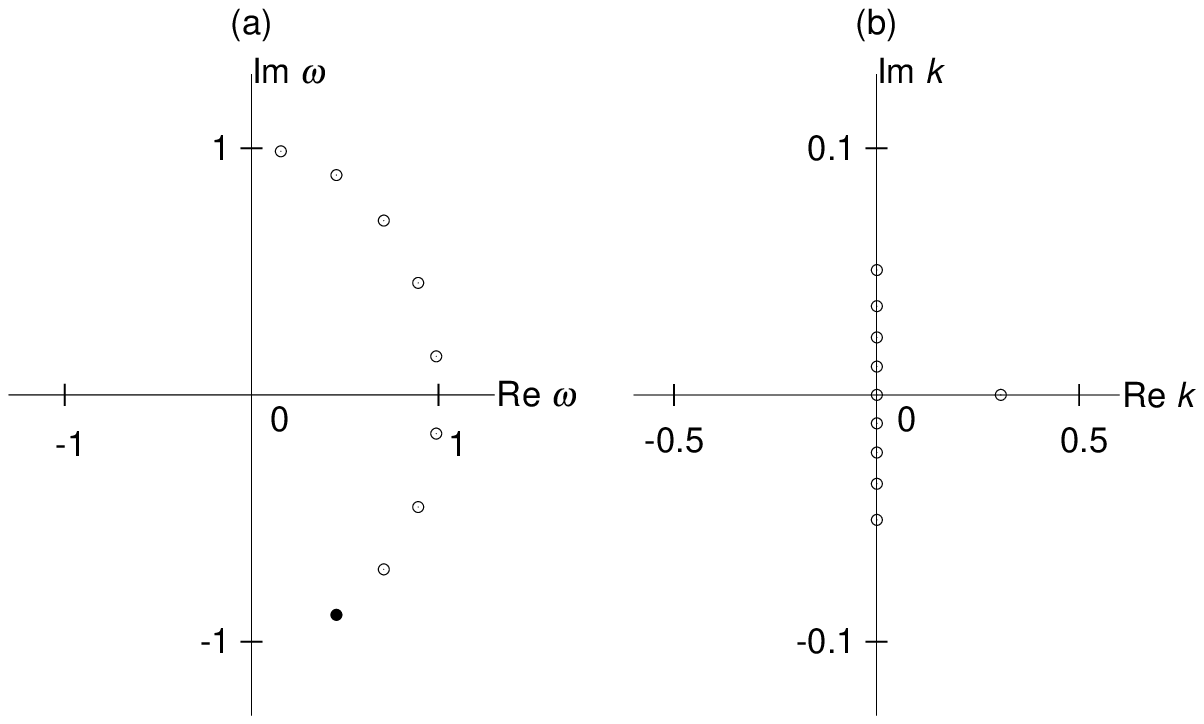}}
\caption{
For the first excited state, (a) the quantum numbers $I_j$'s
are depicted 
in the complex-$\omega$ plane with $\omega=e^{2\pi iI/L}$ and (b) the 
quasi-particle momenta $k_j$'s are shown in the complex-$k$ plane. 
Here, the size of the system $L$ is $20$, the number of particles $n$ 
is $10$,  and the attractive interaction $U$ is $0.0025$. 
The filled circle in (a) is where the two quantum numbers overlap.
} 
\label{momenta}
\efi

\bfi
\centerline{\epsfxsize=9cm   \epsfbox{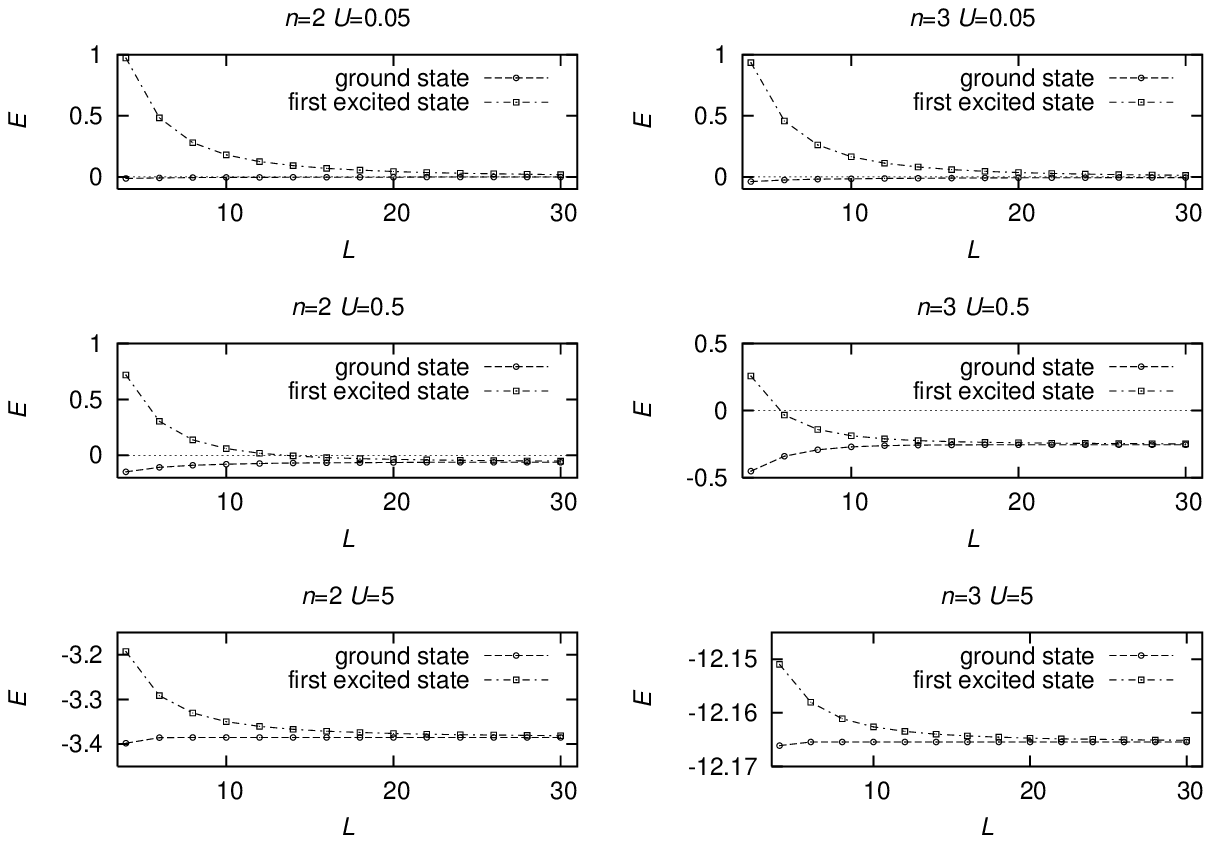}}
\caption{
Ground-state energies and first excited-state energies
are plotted versus the size of the system $L$ ($4 \leq L \leq 30$)
for $U=0.05$, $0.5$, and $5$ in the two- and the three-particle sectors.
The dotted line represents $E=0$.
For all values of $U$ and $L$, the ground-state energy is negative.
On the other hand, when $U=0.5$,
the excited-state energy becomes negative near
$L\simeq 14$ in the two-particle sector and $L\simeq 6$ in the
three-particle sector.
The signs of the excited-state energies for $U=0.05$ and $5$ 
do not change in the range of $L$ shown here.
}
\label{energy}
\efi

\bfi
\centerline{\epsfxsize=9cm \epsfbox{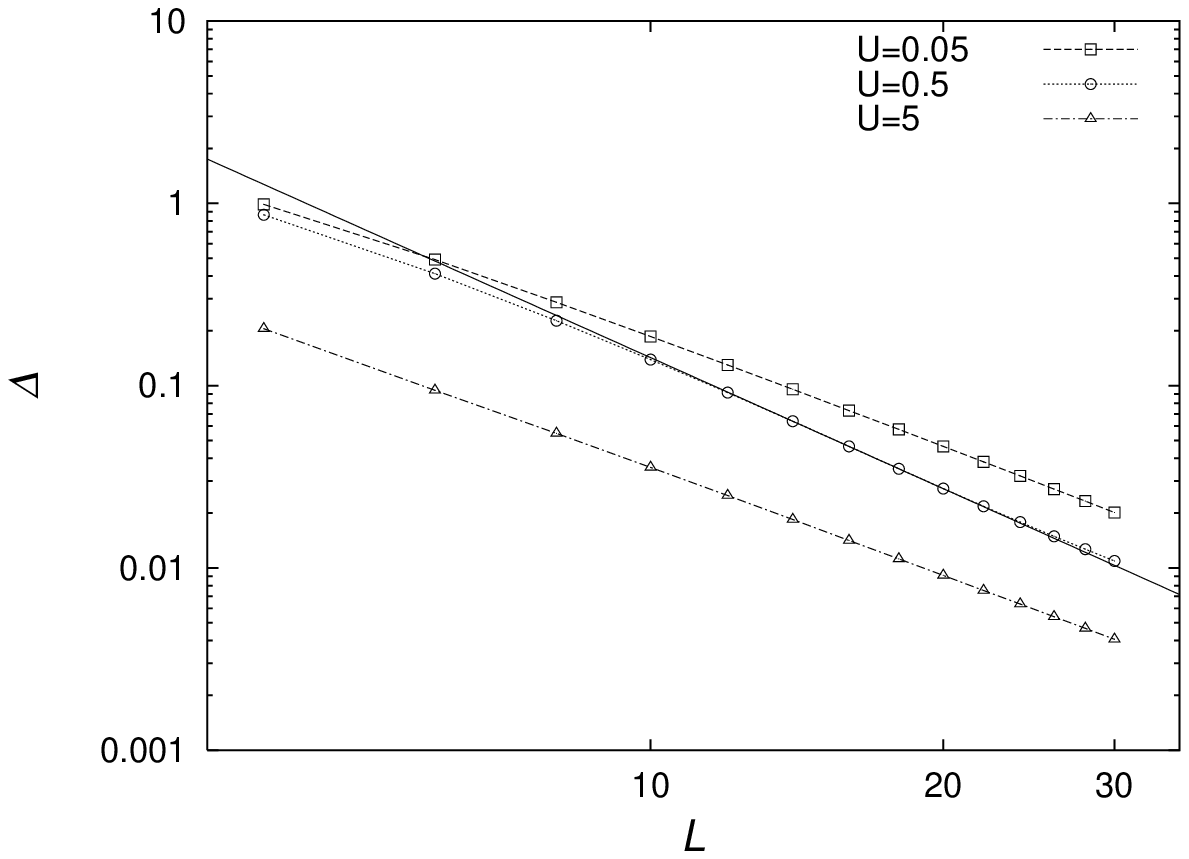}}
\caption{
Log-log plot of the excitation gaps ($\Delta$) versus the size of 
the system ($L$) in the two-particle sector.
Data for $U=0.05$ and $5$ approach  straight lines with slope $z=2.0$, 
but those for $U=0.5$ show a strong crossover before approaching 
the asymptotic behavior. The solid line for $U=0.5$ is that fitted 
in the range $14\leq L\leq 18$, and shows an effective $z\simeq 2.4$. 
}
\label{n=2gap}
\efi

\bfi
\centerline{\epsfxsize=9cm \epsfbox{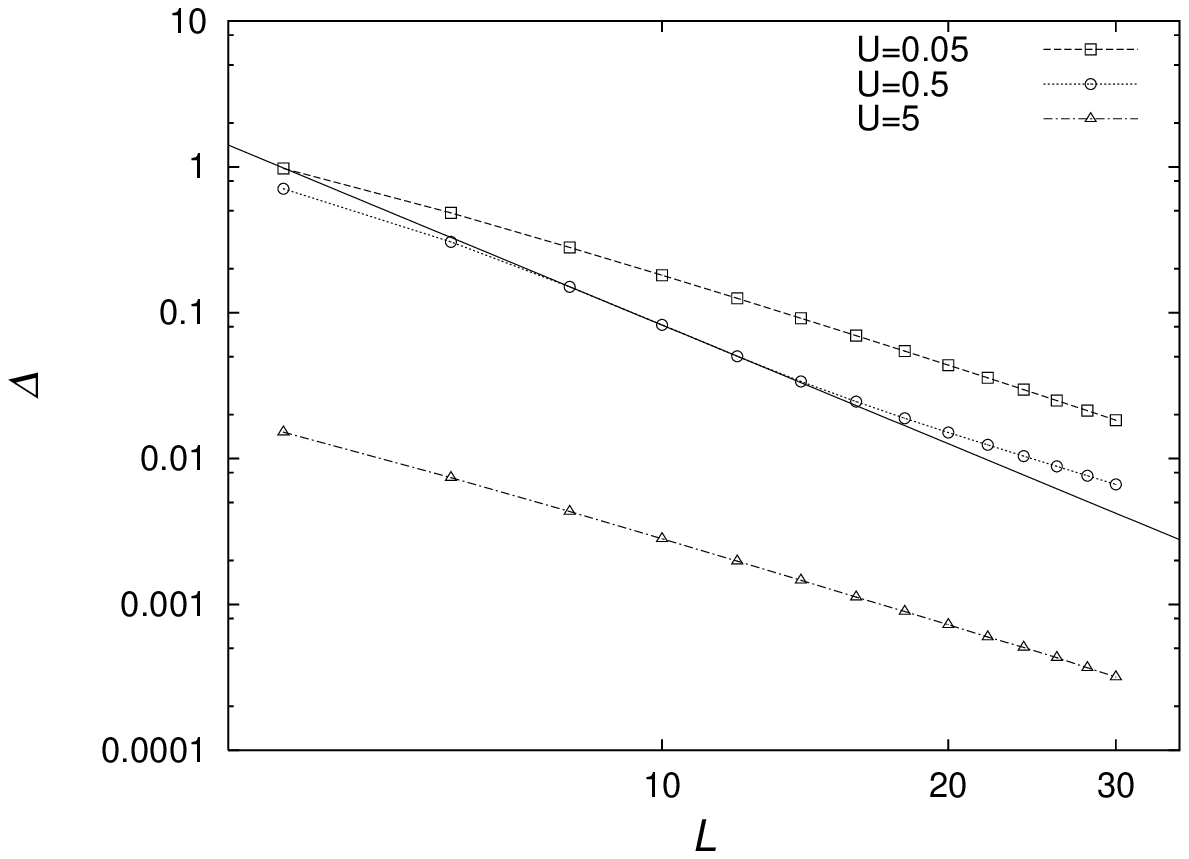}}
\caption{
Same as in Fig.~\ref{n=2gap}, but for the three-particle sector.
The fitted solid line used the data for  $8\leq L\leq 12$, and
has a slope of approximately $2.7$.
}
\label{n=3gap}
\efi

\bfi
\centerline{\epsfxsize=9cm  \epsfbox{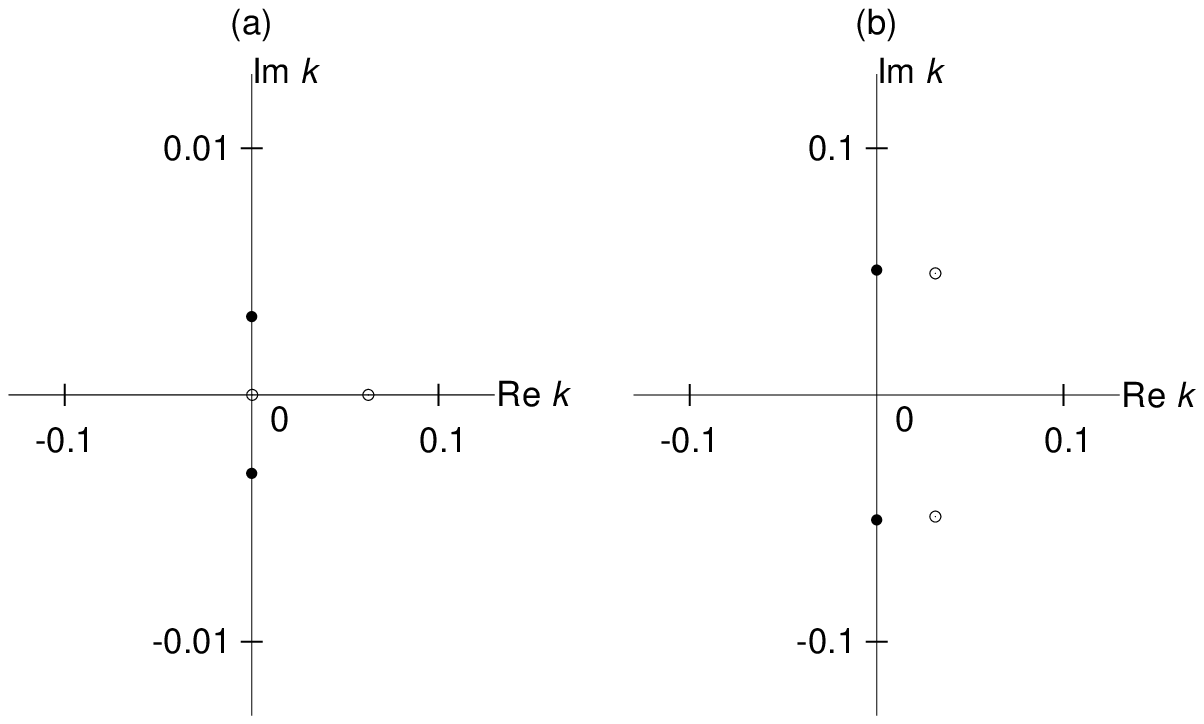}}
\caption{
Distributions of the quasi-particle momenta, $k_j$'s, for the
ground state (filled circles) and the first excited state (open circles) 
are shown in the complex-$k$ plane for $n=2$. 
The size of the system $L$ is $100$ and the interaction $U$ 
is (a) $0.001$ and (b) $0.1$.}
\label{root}
\efi

\bfi
\centerline{\epsfxsize=9cm \epsfbox{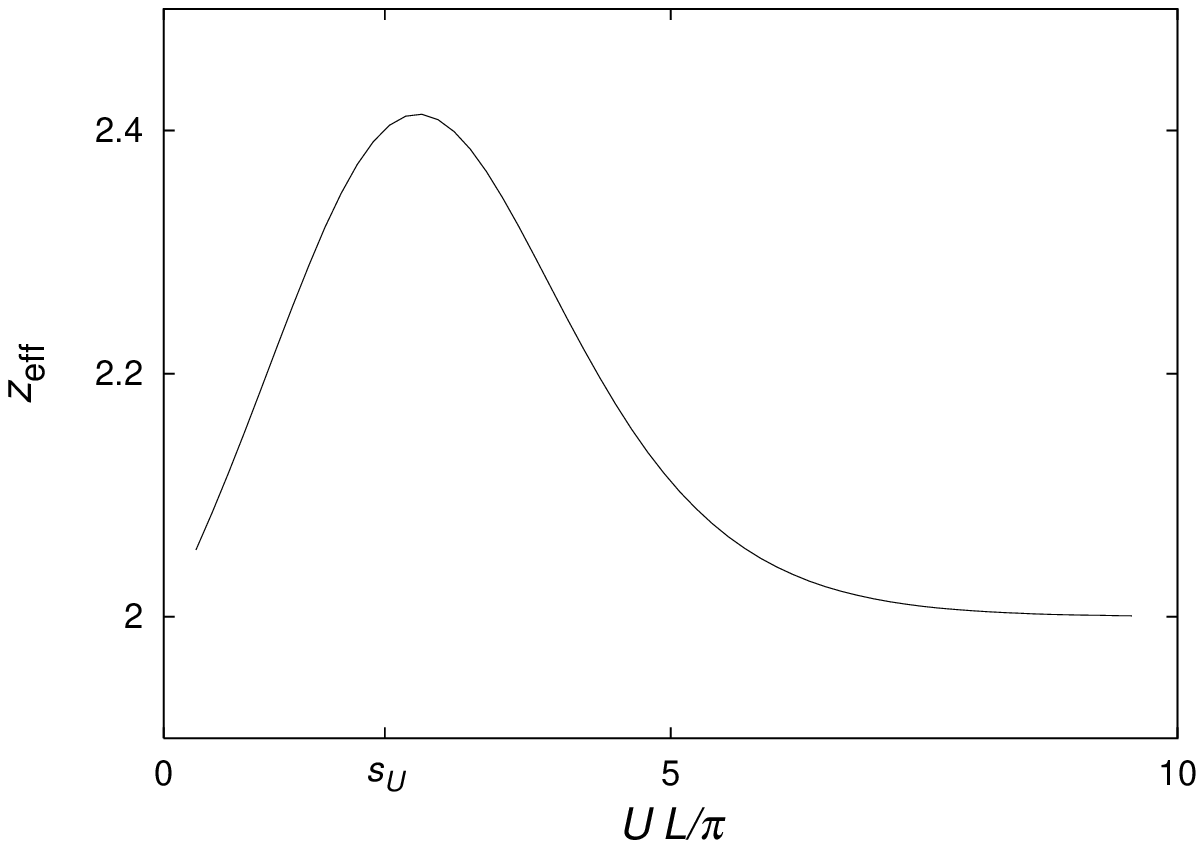}}
\caption{
Effective exponent $z_{\rm eff}$ in the two-particle sector
versus the scaling variable $UL/\pi$ at $L=10000$.
The interaction $U$ varies from $0.0001$ to $0.001$. 
At $UL/\pi=s_U\simeq 2.181$, $z_{\rm eff}\simeq 2.401$.}
\label{z}
\efi 

\end{multicols}


\begin{references}
\bibitem{plischke87}
M. Plischke, Z. Racz, and D. Liu, Phys. Rev. B {\bf 35}, 3485 (1987).
\bibitem{kardar87}
M. Kardar, Nucl. Phys. B {\bf 290} [FS20], 582 (1987). 
\bibitem{spohn92}
L. H. Gwa and H. Spohn, Phys. Rev. A {\bf 46}, 844 (1992).
\bibitem{derrida98}
B. Derrida and J. L. Lebowitz, Phys. Rev. Lett. {\bf 80}, 209 (1998). 
\bibitem{dslee99}
D.-S. Lee and D. Kim, Phys. Rev. E {\bf 59}, 6476 (1999).
\bibitem{appert99}
B. Derrida and C. Appert, J. Stat. Phys. {\bf 94}, 1 (1999).
\bibitem{krug91}
J. Krug and H. Spohn, in {\it Solids Far from Equilibrium},
edited by C. Godr\'{e}che (Cambridge University Press, Cambridge, 1991),
p. 412.
\bibitem{derrida97}
B. Derrida and K. Mallick, J. Phys. A {\bf 30}, 1031 (1997).
\bibitem{park01}
S.-C. Park, J.-M. Park, and D. Kim, unpublished.
\bibitem{brunet00}
E. Brunet and B. Derrida, Phys. Rev. E {\bf 61}, 6789 (2000).
\bibitem{dkim95}
D. Kim, Phys. Rev. E {\bf 52}, 3512 (1995).
\bibitem{henkel94}
M. Henkel and G. Sch\"{u}tz, Physica A {\bf 206}, 187 (1994).
\bibitem{haldane82}
T. C. Choy and F. D. M. Haldane, Phys. Lett. {\bf 90A}, 83 (1982).
\bibitem{lieb63}
E. H. Lieb and W. Liniger, Phys. Rev. {\bf 130}, 1605 (1963).
\bibitem{Muga98}
J. G. Muga and R. F. Snider, Phys. Rev. A {\bf 57}, 3317 (1998).
\bibitem{krauth91}
W. Krauth, Phys. Rev. B {\bf 44}, 9772 (1991).
\end{references}
\end{document}